# How do machines learn?

## Evaluating the AIcon2abs method

**Rubens Lacerda Queiroz**[1] · **Cabral Lima**[1,2] · **Fábio Ferrentini Sampaio**[3] · **Priscila Machado Vieira Lima**[4,5]

1 PPGI, Federal University of Rio de Janeiro, Rio de Janeiro, Brazil
2 Computer Science Institute, Federal University of Rio de Janeiro, Rio de Janeiro, Brazil
3 Polytechnic University of Setúbal – Portugal
4 PESC/COPPE, Federal University of Rio de Janeiro, Rio de Janeiro, Brazil
5 Tercio Pacitti Institute (NCE), Federal University of Rio de Janeiro, Rio de Janeiro, Brazil

### Abstract

This study expands on previous work that introduced the AIcon2abs method (AI from Concrete to Abstract: Demystifying Artificial Intelligence to the general public), an innovative approach designed to increase public understanding of machine learning (ML) across diverse age groups, including K–12 students, and aims to evaluate its effectiveness. AIcon2Abs employs the WiSARD algorithm, a weightless neural network known for its simplicity, and user accessibility. WiSARD does not require Internet, making it ideal for non-technical users and resource-limited environments. This method enables participants to intuitively visualize and interact with ML processes through engaging, hands-on activities, as if they were the algorithms themselves. The method allows users to intuitively visualize and understand the internal processes of training and classification through practical activities. As WiSARD operates without Internet connectivity, it can learn effectively from a minimal dataset, even from a single example. This feature enables users to observe how the machine improves its accuracy incrementally as it receives more data. Moreover, WiSARD generates *mental images* representing what it has learned, highlighting essential features of the classified data. AIcon2abs was tested through a six-hour remote course with 34 Brazilian participants, including 5 children (8–11 yo), 5 adolescents (12–17 yo), and 24 adults (21–72 yo). Data analysis was conducted from two perspectives: a mixed-method pre-experiment (including hypothesis testing), and a qualitative phenomenological analysis. Nearly 100% of participants rated AIcon2abs positively, with the results demonstrating a high degree of satisfaction in achieving the intended outcomes. This research was approved by the CEP/HUCFF/UFRJ Research Ethics Committee.

## 1. Introduction

As noted by Queiroz et al. (2021), Artificial Intelligence (AI) has been increasingly integrated into a wide variety of fields, influencing numerous facets of daily life at an ever-expanding rate. In response to this rapid growth, several scholars have highlighted the critical need to develop a strategic framework for the advancement of AI, stressing that such a strategy must emerge from a reflective process that involves all sectors of society (Almeida, 2018; Caputo, 2019; Medina, 2004; Yogeshwar, 2018).

Nonetheless, a pivotal question arises: How can individuals effectively participate in discussions and decision-making regarding the future of AI without possessing a foundational understanding of the field? This dilemma underscores an urgent necessity to develop accessible methods that enable the general public, including K–12 students, to become informed and engaged participants in debates and decisions concerning the adoption and implementation of AI technologies. A major challenge in achieving this goal lies in the inherent complexity of the techniques and concepts underpinning most AI systems (Sakulkueakulsuk et al., 2018). In response, researchers have begun to explore solutions aimed at overcoming these obstacles and fostering a more accessible understanding of AI.

Russell and Norvig (2020) classify the AI domain into three principal areas: Search, Knowledge Representation, and Machine Learning. Among these, Machine Learning has emerged as one of the most significant fields, largely responsible for the recent "AI boom." Most AI systems that individuals encounter in their daily lives are built on Machine Learning principles, as in the case of Generative AI[1] as chat GPT[2] and other LLMs[3], which combines machine learning mechanisms with other AI approaches to produce generative AI[4]. Consequently, most initiatives focused on demystifying AI for the public have concentrated on the Machine Learning component, as evidenced by the Multivocal Literature Review (MLR)[5] conducted by Queiroz et al. (2021). This review encompassed scientific databases and general internet searches, leading to the identification of various artifacts, approaches, characteristics, and unresolved issues. From these findings, the AIcon2abs method was developed as a potential solution (Queiroz et al., 2021).

The AIcon2abs method was designed to facilitate an introductory understanding of AI for the general public, using Machine Learning as an accessible entry point. *In its most basic form, Machine Learning involves providing an algorithm with examples of a specific class, which enables the system to later classify new observations as belonging or not to that class*. For the purposes of this study, the foundational understanding of Machine Learning is framed by two core concepts: ***(i) Machine Learning is essentially a process by which computers learn from examples, and (ii) this learning occurs through the execution of a computer program.***

The AIcon2abs method comprises several key components, including a block-based programming environment, a series of interactive educational activities based on the WiSARD weightless artificial neural network model, a set of educational robotics materials, and a standard instructional unit (IU). The block-based programming environment, termed **BlockWiSARD** (Queiroz et al., 2021), was specifically designed to support individuals with no prior programming experience in (i) creating programs that enable their computers to learn how to recognize images and (ii) understanding the underlying operations of these programs. Notably, BlockWiSARD can be operated on low-cost computers without the need for a GPU or internet connection.

In BlockWiSARD, higher-level tasks necessary for machine learning, as well as the instantiation of the learning algorithm, are incorporated as commands within the program being developed. This feature is intended to facilitate a clearer distinction between traditional computer programs and those capable of learning from examples. Specifically, BlockWiSARD includes four programming blocks directly related to the training and classification process. By combining these blocks with conventional programming commands, users can easily construct systems that (i) instantiate a learning mechanism, (ii) input data for learning, (iii) assign corresponding labels to that data, and (iv) classify new input data.

---

[1] Simply put, generative AI generates content, which could be text, images, or multimedia.

[2] Chat GPT is an AI Large Language Model (LLM) developed by Open AI to follow instructions in a prompt and provide detailed responses. *https://openai.com/blog/chatgpt*

[3] "LLMs (Large Language Models) are a type of artificial intelligence (AI) that can generate, understand, and interact with natural language, such as text and speech. Natural language is the way we humans communicate with each other, using words, sentences and grammar" (Dash, 2023, p.1)

[4] "Generative AI is a form of artificial intelligence that utilizes techniques from machine learning and deep learning to generate original data" (Kalota, 2024)

[5] A Multivocal Literature Review (MLR) comprises all accessible writings on a usual, often contemporary topic (Ogawa & Malen, 1991).

The concept of incorporating machine learning primitives into programming languages was originally proposed by Tom Mitchell (2006). In addition to the machine learning blocks, BlockWiSARD features a block responsible for generating a *mental image* of a given class. These mental images enable users to visualize a prototype of a class learned from training patterns, thereby enhancing their understanding of how machine learning models conceptualize and categorize data.

An essential principle of BlockWiSARD is that each command block within the program must correspond to a tangible action in the concrete world, such as displaying an image on a screen, hearing a sound played a specific number of times, or seeing a light blink. This premise must also be valid for the machine learning related programming blocks. But how can we visualize or observe the internal processes carried out by the machine when the learning algorithm is instantiated? What happens when the Learn and Recognize commands are executed in the program? Additionally, how can we enable individuals to replicate these actions as if they were the algorithm themselves? To this end, the AIcon2abs method incorporates games, animations, and interactive activities centered around the WiSARD model, designed to engage users in a hands-on, experiential learning process (Queiroz et al., 2021; Queiroz, 2023).

As Papert (1994, p.31) said, "The reason you are not a mathematician might well be that you think that math has nothing to do with the body; you have kept your body out of it because it is supposed to be abstract, or perhaps a teacher scolded you for using your fingers to add numbers! This idea [...] has inspired me to use the computer as a medium to allow children to put their bodies back into their mathematics." WiSARD is a perfect tool to put people's bodies into learning AI through ludic activities. These activities also enable learners to observe the internal operations of the machine when the learning algorithm is instantiated within BlockWiSARD. They also allow the visualization of what the machine does when the Learn and Recognize blocks are used in the BlockWiSARD programs. Moreover, the ludic activities make it possible for people to perform all those actions as if they were the WiSARD algorithm itself.

BlockWiSARD operates in three modes: (i) Computer Only; (ii) Computer + Arduino[6], or (iii) Raspberry Pi[7]. The latter two modes enable the integration of BlockWiSARD with low-cost educational robotics platforms. In addition to the core materials, a set of robotics tools compatible with BlockWiSARD was developed as part of the AIcon2abs method (Queiroz et al., 2021). While not essential to the method, these materials complement the learning process and demonstrate how robotics can enhance the educational experience when combined with BlockWiSARD.

For the effective implementation of the AIcon2abs method, the inclusion of a standard instructional unit (IU) is essential. This IU includes an asynchronous e-learning implementation with a workload of approximately 6 hours which was used to evaluate the method, as discussed in Section 3 of this paper. ***This evaluation is the focus of this paper.***

The theoretical foundations supporting the design of the AIcon2abs method are rooted in several core concepts and technologies, including: (i) Piaget's (1950) Constructivist Theory, (ii) Papert's (1993) Constructionist Theory, (iii) Russell and Norvig's (2020) concept of Intelligent Agents, (iv) the Machine Learning process proposed by Mitchell (1997), (v) Turing's (1950)

---

[6] Arduino is a prototyping platform simple enough to be used by any student, including K–12 students. The board's design allows people to "build things that work" without attending a complete electronics course (Banzi, 2012; Severance, 2013). Its low-cost, quality, flexibility, and ease of use have made Arduino electronic prototyping boards an excellent option for developing projects on Educational Robotics *http://www.arduino.cc/*

[7] Raspberry Pi is a low-cost single-board computer provided with a set of General-Purpose Input/Output pins where one can connect robotics devices. Roughly speaking, we can say that the Raspberry Pi is a single-board computer with a microcontroller (like an Arduino) built into it. The flexibility of the Raspberry Pi allows one to use it for the most diverse purposes, such as experimentation, residential automation, or simply fun. (Severance, 2013). *https://www.raspberrypi.org/*

Imitation Game, (vi) Feynman's (1989) notion of learning through construction, (vii) Block-based programming (Resnick et al., 2009), (viii) the WiSARD weightless artificial network (Aleksander et al., 2009), (ix) Active Learning methodologies (Bonwell & Eison, 1991), and (x) educational robotics (Eguchi, 2010; Papert, 1994). These concepts and their interconnections with the AIcon2abs method are explored in further detail by Queiroz et al. (2021) and Queiroz (2023).

## 1.1 Research question

Queiroz et al. (2021) provide a comprehensive theoretical framework for the conceptualization of the AIcon2abs method and translate these theoretical ideas into functional artifacts with significant potential to address the problem the method aims to solve. *However, in that first study, the authors did not conduct an empirical evaluation of the method*. In light of this, the present study seeks to empirically evaluate the AIcon2abs method in order to answer the following research question: ***Is AIcon2abs a suitable method to support a diverse audience of different ages and backgrounds in understanding the concept of machine learning in its most basic form in an uncomplicated and satisfactory way?***

## 1.2. Objectives

### 1.2.1. General objective

The primary aim of this study is to evaluate the AIcon2abs method in order to assess its potential strengths and weaknesses as a tool for helping an audience of eight years old and over to understand satisfactorily, in an uncomplicated way, the concept of machine learning in its most basic form. The concept of *satisfactory solution*, as defined by Simon (1996), refers to a solution that adequately addresses a given problem. In this context, a satisfactory solution may be achieved in two ways: (i) through consensus among the stakeholders involved in the problem or (ii) by advancing the current solution relative to previous solutions provided by earlier artifacts (Dresch et al., 2015).

### 1.2.2. Specific objectives

The specific objectives of this research are as follows: (i) to evaluate the artifacts developed as part of the AIcon2abs method, ensuring they meet the specified requirements; (ii) to assess the validity of the theoretical conjectures formulated in Subsection 1.4; and (iii) to formalize the construction and contingency heuristics of the AIcon2abs method (see Subsections 4.3.4 and 4.3.5), which constitute the primary research contributions outlined by the Design Science Research (DSR) methodology proposed by Dresch et al. (2015) (Section 2).

## 1.3. AIcon2abs method requirements

The AIcon2abs method was developed with the intention of fulfilling the following requirements (Queiroz et al., 2021):

1. It must be suitable for individuals both with and without fully developed abstraction capabilities.
2. It must incorporate a machine learning algorithm that facilitates the easy visualization and manipulation of its underlying logic through interactive visual elements.

3. It must provide a block-based programming environment that enables non-expert users (including K–12 students) to create basic programs capable of learning. *The program environment must include the chosen learning algorithm's instantiation, training, and classification commands among its primitives*, and it must be compatible with low-cost robotic platforms.
4. The method should adopt Active Learning methodologies, including narratives, challenges, interactive activities, and games.
5. *It must be designed to operate effectively on low-cost computers which do not require a Graphics Processing Unit (GPU) or an internet connection.*

### 1.4. AIcon2abs theoretical conjectures

*Theoretical conjectures* represent assumptions regarding human or social behavior, grounded in theories, models, and constructs from various disciplines. Within the framework of Design Science Research (DSR), these conjectures guide the design of the artifact, and the artifact is then employed to test the validity of these conjectures (Pimentel et al., 2019). The core theoretical conjecture that informed the design and development of the AIcon2abs method is as follows:

1. ***Learning abstract concepts from computer science can be facilitated through practices that highlight observable aspects of these concepts in the concrete world.***

Additional theoretical conjectures that contributed to the design of the AIcon2abs method include:

2. Algorithms with explicit visual representations enhance understanding for non-expert audiences, including K–12 students.
3. A machine learning algorithm capable of online learning (interleaving training and classification) and generalizing from only one or a few trained examples facilitates the understanding of the machine learning process described by Mitchell (1997).
4. General users, including K–12 students, can grasp the distinction between conventional computer programs and those capable of learning by: (i) integrating the instantiation of the machine learning algorithm, along with training and classification primitives, into basic programming constructs; (ii) adopting an agent-based approach; and (iii) utilizing the machine learning process framework articulated by Mitchell (1997).
5. The essence of the Imitation Game (Turing, 1950) provides a valuable tool for comprehending AI, particularly by comparing the performance of a human and a machine in tasks requiring human-like intelligence.

### 1.5. Expected research subjects’ performance

By adhering to the theoretical conjectures and fulfilling the outlined requirements, the AIcon2abs method is expected to enable its target audience to achieve the following outcomes:

1. Develop the ability to create programs utilizing basic programming structures such as loops and conditionals.
2. *Replicate, through ludic activities, the specific steps performed by the machine learning algorithm used in the programs they develop.*
3. ***Understand the fundamental process of teaching a machine to learn from examples using machine learning.***

## 2. Research methodology and text structure

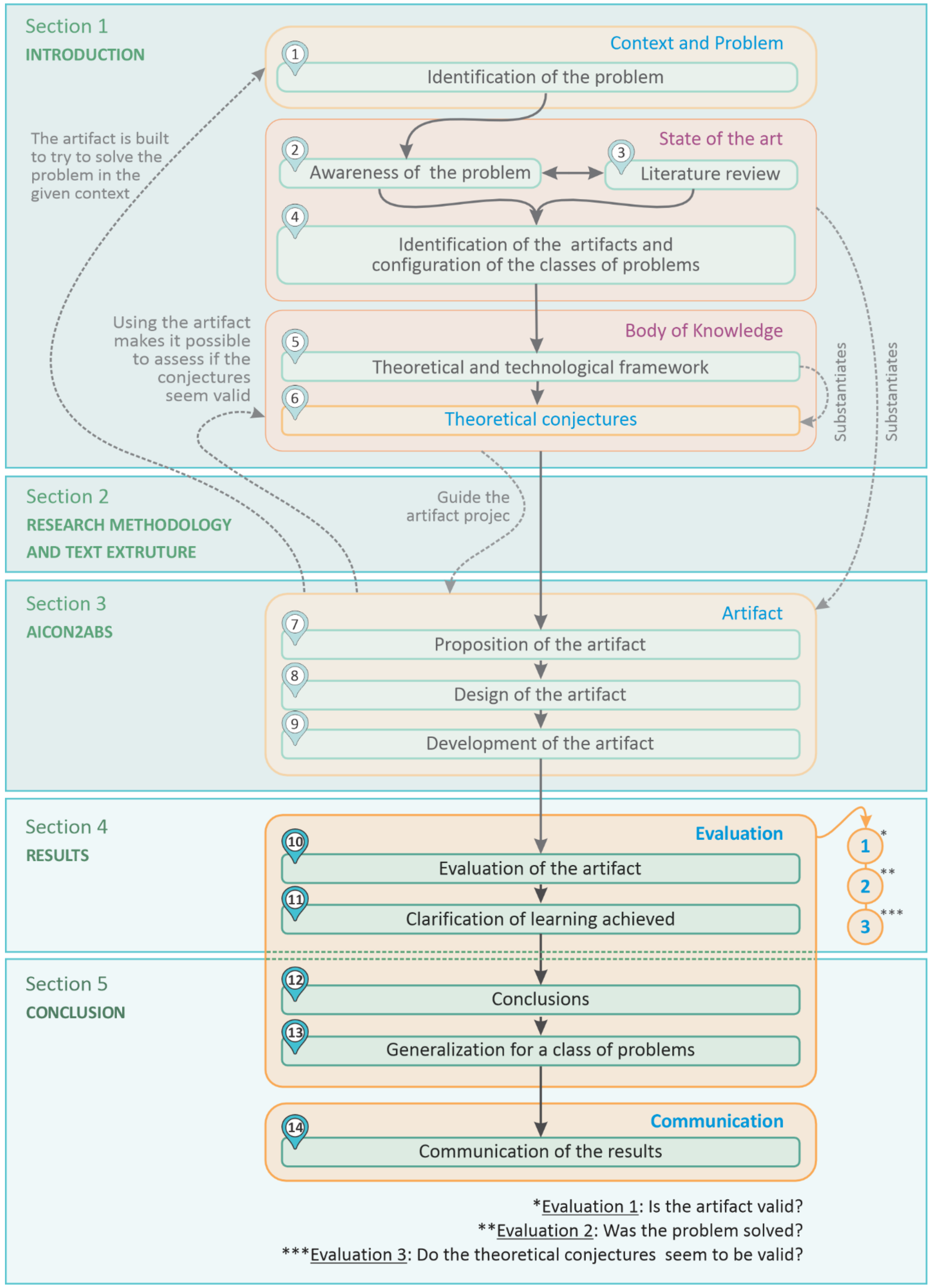

**Fig. 1** DSR steps and text structure. Adapted from Dresch et al. (2015) and Pimentel et al. (2019)

To address our research question, we employed the Design Science Research (DSR) method outlined by Dresch et al. (2015), incorporating adaptations from the DSR map elements developed by Pimentel et al. (2019). Notably, the DSR map by Pimentel et al. (2019) does not constitute a research method per se; rather, it serves as a framework to support DSR-oriented investigations, irrespective of the specific DSR method adopted. By integrating these approaches, we established a process comprising 14 steps organized into six clusters (see Figure 1). This article is structured in alignment with the DSR steps depicted in the diagram presented in Fig. 1, with a particular focus on the evaluation of the AIcon2abs method and the communication of findings, covering steps 10 through 14.

Section 1 provides the context for this study, presenting an overview of the AIcon2abs method's conception, structure, and the objectives of this research. Section 2 delineates the research methodology, followed by Section 3, which details the design, development, and implementation of the AIcon2abs standardized user interface (UI). Sections 4 and 5 then concentrate on the evaluation of the AIcon2abs method. A comprehensive discussion of steps 1 through 9 can be found in Queiroz et al. (2021) and Queiroz (2023), the latter being Queiroz's Doctoral Thesis.

Dresch et al. (2015) suggest that DSR-based research may integrate data collection and analysis techniques from other methodological frameworks, such as case studies and experimental designs. In this study, we adopted a pre-experimental design alongside a phenomenological approach to data collection and analysis. A pre-experiment involves limited control over the variables involved in the process, employing a single group (the experimental group) where the intervention is implemented without a control group (Cohen et al., 2002). For the phenomenological component, we applied the approach suggested by Creswell (2016), which frames phenomenology as a qualitative method to explore the shared meanings individuals ascribe to their lived experiences regarding a specific phenomenon (Van Manen, 1990).

## 3. AIcon2abs instructional unit

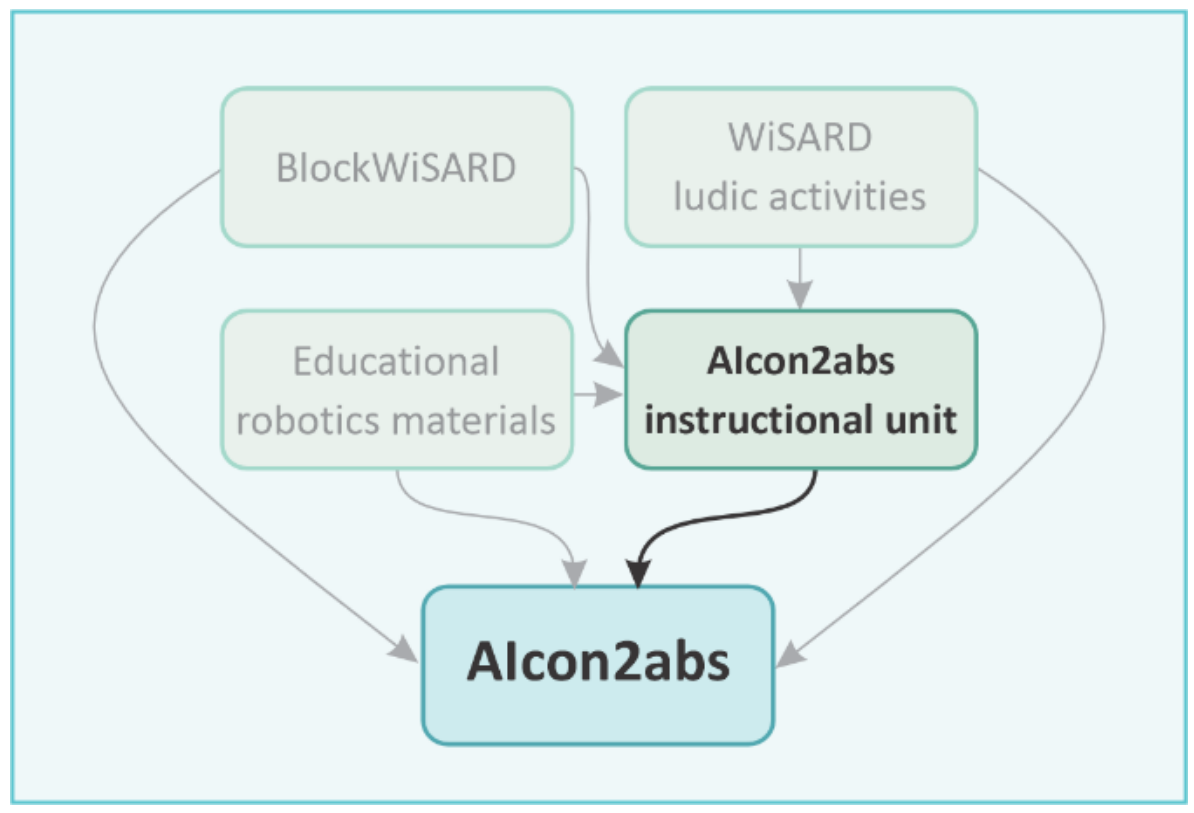

**Fig. 2** AIcon2abs components' method summary

Figure 2 illustrates a comprehensive diagram of the AIcon2abs method's components. The standardized AIcon2abs interface (IU) was implemented as the primary tool in the empirical study to rigorously assess and validate the method. For additional information on the BlockWiSARD block-based programming environment, Ludic Activities related to the WiSARD Weightless Artificial Neural Network (WANN), and AIcon2abs Educational Robotics Materials, refer to Queiroz et al. (2021).

### 3.1. AIcon2abs IU components

The development of the AIcon2abs IU adhered to the ADDIE instructional design model (Analyze, Design, Develop, Implement, and Evaluate) as advocated by Branch (2009) and Schlegel (2012). Among various instructional design frameworks, ADDIE is recognized for its efficacy in addressing complex educational scenarios, making it well-suited for the creation of comprehensive learning resources (Branch, 2009).

To facilitate this empirical evaluation, forty-one video lessons and fourteen interactive ludic activities were developed in Portuguese for the AIcon2abs IU. These learning resources were organized into seven instructional units as follows: (1) Introduction; (2) Introducing the Story of Zrow; (3) Transforming Your Computer into a Zrow Prototype; (4) Exploring BlockWiSARD – Zrow's Programming Environment; (5) Teaching Zrow about Star and Flower Shapes; (6) WiSARD – The Learning Engine of Zrow; and (7) Zrow's Departure. All AIcon2abs reference materials (in Portuguese) are available in Queiroz (2023). The Portuguese video lessons and ludic activities can also be accessed via the following YouTube[8] playlists:

1. Invitation to the AIcon2abs Course (https://www.youtube.com/playlist?list=PLGOt20Ava9YDbXosELwFw_tv3CHL4kkOg)
2. AI for the General Public: Introduction (https://www.youtube.com/playlist?list=PLGOt20Ava9YCHZaaYQeVq2D8TiasBnwc-)
3. AI for the General Public: Introducing the Story of Zrow (https://www.youtube.com/playlist?list=PLGOt20Ava9YCjFb3Of5NrZdN2PJV1SWn6)
4. AIcon2abs Course Videos (https://www.youtube.com/playlist?list=PLGOt20Ava9YB-nlAZgqpeuHp53uQLTE7x)

Examples of interactive activities in English are also available at:

1. The Digitized Game (https://scratch.mit.edu/projects/453213702)
2. You Are a WiSARD! (https://scratch.mit.edu/projects/440606183)
3. You Are a WiSARD with Bleaching! (https://scratch.mit.edu/projects/440606183)

The instructional design included the formation of three Google Classroom groups segmented by age: children (8–12 years), adolescents (12–17 years), and adults (18+ years), in compliance with the Brazilian National Research Ethics Committee (Comissão Nacional de Ética em Pesquisa – CONEP[9]) guidelines. All groups accessed the same learning content and environment. Prior to deployment, navigability testing was conducted: all video lessons were viewed, and all activities were completed to identify and rectify any potential issues.

To promote the course, materials were created in three formats—text, image, and video—and disseminated via YouTube, Facebook[10], WhatsApp[11], and Instagram[12]. Video-based student guides were also developed, providing instructions on course progression and usage of the virtual learning platform. **Pre-test and post-test questionnaires** were prepared and diostributed through Google Forms.

---

[8] *https://www.youtube.com/*
[9] *http://conselho.saude.gov.br/comissoes-cns/conep/*
[10] *https://www.facebook.com/*
[11] *https://www.whatsapp.com/*
[12] *https://www.instagram.com/*

### 3.2. AIcon2abs IU implementation

Following the development and testing of the AIcon2abs standardized IU, the instructional materials created for this method were delivered through an online course titled "AIcon2abs" (*AI from Concrete to Abstract, Demystifying Artificial Intelligence to the General Public*). Details of this process are provided in Subsections 3.2.1 and 3.2.2. To access the AIcon2abs course for the method evaluation, users can follow the steps below using a Google account:

1. **Use the invitation link:** https://classroom.google.com/c/NjE3NDgyNjEwMzE3?cjc=6bndekk

   or

1. Access *Google Classroom*.
2. Click on the (+) icon in the upper corner of the page.
3. Choose the option *Join Class*.
4. Enter the class code **6bndekk** and click *Join*.
5. Finally, click on the classwork tab.

#### 3.1.1. Initiation, conduction, and completion of instruction

Participants had the flexibility to engage with the course activities at their convenience, allowing for asynchronous participation and independent task completion. This design permitted participants to join the course at different times and progress at their own pace. Any questions or feedback from participants were managed through messages within the platform.

**The AIcon2abs course was completed by thirty-four participants aged 8 to 72**. This cohort included twenty-four adults (aged 21–72), five adolescents (aged 12–17), and five children (aged 8–11). Of the participants, 53% had previous programming experience, while 23% had prior knowledge of AI or machine learning. Participants' educational backgrounds were diverse, with 32% in Social, Biological, and Human Sciences, 38% in the exact sciences and engineering, and 29% in basic education. Geographically, 47% resided in Southeast Brazil, and 53% in the Southern region of the country.

#### 3.1.2. Data collection instruments

The primary instrument for data collection was a set of **pre-test and post-test questionnaires** comprising both open-ended and closed questions. Closed questions included nominal (yes/no and multiple-choice) and ordinal (Likert scale) formats. The pre-test questionnaire contained 11 items, while the post-test questionnaire consisted of 15 items. These questionnaires are available in Queiroz (2023). Secondary data, including public posts, private messages within the learning platform, and email communications, also contributed to the data collection process.

## 4. Results

As outlined in Section 2, data from the pre-test and post-test questionnaires, along with secondary data sources, were analyzed from two perspectives: (i) **as part of a mixed-methods pre-experiment** (**including exploratory data analysis and hypothesis testing**) and (ii) **from a phenomenological perspective** (a qualitative examination capturing the essence of participants' experiences with the AIcon2abs course).

This section presents findings related to participants' comprehension of the machine learning process and their evaluation of the usability and clarity of the AIcon2abs method components (see Table 1). The questions used in this analysis, available in Queiroz (2023), focused on participants' basic understanding of machine learning (**in its most basic form**) before and after the course, as well as their insights regarding the course materials and their overall experience in the AIcon2abs course.

**Table 1** List of questions used for the data analysis presented in this article

| Kind | ID | Question | Data type |
|---|---|---|---|
| Pre-post | Q03 | If you believe these devices can learn new things, tell me how you think they are taught | Categorized |
| Pre-post | Q04 | If you believe that the computer can recognize an image it has never "seen" but is similar to other photos it has "seen," tell me: how do you think it can do that? | Categorized |
| Pre-post | Q07 | What does it take to "transform" a conventional computer program into a program capable of learning? | Categorized |
| Pre-post | Q10 | What is machine learning? | Categorized |
| Post | Q14 | Has your understanding of machine learning changed after you took the course? | Categorical |
| Post | Q15 | What has changed? | Open |
| Post | Q18 | Did you understand the programs made with BlockWiSARD? | Ordinal |
| Post | Q20 | Did you understand how WiSARD works by doing the Scratch[13] [interactive ludic] activities? | Ordinal |
| Post | Q23 | Did you like to take the AIcon2abs course? Were you satisfied with the things you learned during the process? | Ordinal |
| Post | Q25 | What did you learn by taking this course? | Open |
| Post | Q26 | Did the course bring you any reflections? Which? | Open |

For this analysis, we employed two types of categorical data: (i) **inherently categorical data and** (ii) **derived categorical data**. Inherently categorical data were obtained from responses to questions with predefined options, such as Yes/No and multiple-choice answers. Derived categorical data, on the other hand, were generated by categorizing responses to open-ended questions. To avoid influencing participants' responses, some questions were intentionally left open-ended without predefined options. Categorization was applied when responses could be grouped by thematic similarity, enabling both visualization and quantification of the data. Consequently, wherever feasible, answers to open-ended questions were systematically categorized.

Table 2 presents examples of categorized responses to the question: "*If you believe these devices can learn new things, explain how you think they are taught*." A comprehensive categorization of all open-ended responses is available in AIcon2abs (2023). The categorization and scoring process for these responses adhered to the methodologies proposed by Smith (2000) and Woike (2007). Categorical and categorized data visualization and analysis are presented for the full participant dataset, with certain visualizations organized by age group where relevant. Since the analyzed data are exclusively categorical or ordinal, calculations and visualization tools appropriate solely for interval and ratio data in exploratory analysis were not employed in this study.

[13] https://scratch.mit.edu/

**Table 2** Examples of categorization of open-ended answers

| Original answer | Categorization |
|---|---|
| *They are taught through the presentation of some examples of a particular class so that, later, reading, and interpreting patterns.* | With examples |
| *I'm not sure, but I think these devices learn from training. For example, a robot can be "trained" to recognize facial expressions.* | Trained by humans |
| *They are taught with programs and images.* | Computer programs |
| *With data provided by different users over time.* | Collecting data |
| *Machine Learning.* | Machine learning |
| *Do not learn.* | Do not learn |

In the case of pre- and post-test questions, a modified application and interpretation of the confusion matrix was utilized as a data visualization tool, which we termed the **Pre-Posttest Matrix**. Originating with Pearson (1904), the confusion matrix, or contingency table, is defined as follows: "A confusion matrix for an N-way classification task is an N by N matrix where the cell (x, y) contains the number of times an item with correct classification x was classified by the model as y" (Jurafsky & Martin, 2008, p.35). Correspondingly, **a Pre-posttest Matrix for an *N*-possible-answers pretest-posttest question *Q* is an *N* by N matrix where the cell (*x*, *y*) contains the number of times a question *Q* answer changed its value from *x* in the pre-test to *y* in the post-test. The Pre-Posttest Matrix is designed to illustrate more explicitly the shifts in participants' responses to a given question between the pre-test and post-test**.

It is also noteworthy that, prior to the implementation of the AIcon2abs course, a beta test[14] was conducted to evaluate the ludic activities centered around the WiSARD model. This test was administered to undergraduate students enrolled in an introductory computer programming course. After engaging in these activities, the students successfully implemented fully functional WiSARDs based solely on the knowledge they acquired from the exercises. All participants rated the activities involving the WiSARD model as accessible and engaging. Informed consent was obtained from all students prior to their participation in the beta test.

### 4.1. Data analysis from a pre-experiment perspective

This section provides an exploratory analysis and **hypothesis testing** using a mixed-methods approach, based on data collected and analyzed within a pre-experimental design framework. For the purpose of this analysis, the adult participant group was subdivided into two distinct age categories: (i) Adults, comprising individuals aged 21 to 59, and (ii) Seniors, comprising individuals aged 60 to 72. This age-based division aligns with the United Nations guidelines[15].

[14] A type of user testing that uses testers who are NOT part of your organization and who are members of your product's target market. The product under test is typically very close to completion." (kaner et al., 2008, p.34)
[15] https://www.un.org/development/desa/pd/sites/www.un.org.development.desa.pd/files/unpd_egm_201902_s1_sergeischerbov.pdf

### 4.1.1. **Pre-posttest questions exploratory analysis**

***Q03-Pre-posttest:*** *If you believe these devices can learn new things, tell me how you think they are taught. (Categorized)*

The bar chart in Figure 3a illustrates that, during the pre-test phase (labeled as Q1), none of the participants expressed the concept of using examples to teach a computer. Additionally, only three participants identified training as a method for teaching a computer. However, in the post-test phase (labeled as Q2), the number of participants who verbalized these concepts increased to 16, indicating an 80% rise. Notably, of the six participants who either did not know or did not respond to the question during the pre-test, only one retained this non-response in the post-test. The stacked bar charts in Figure 3b and Figure 3c further highlight a perceptible improvement in the quality of responses to this question across all age groups from pre-test to post-test.

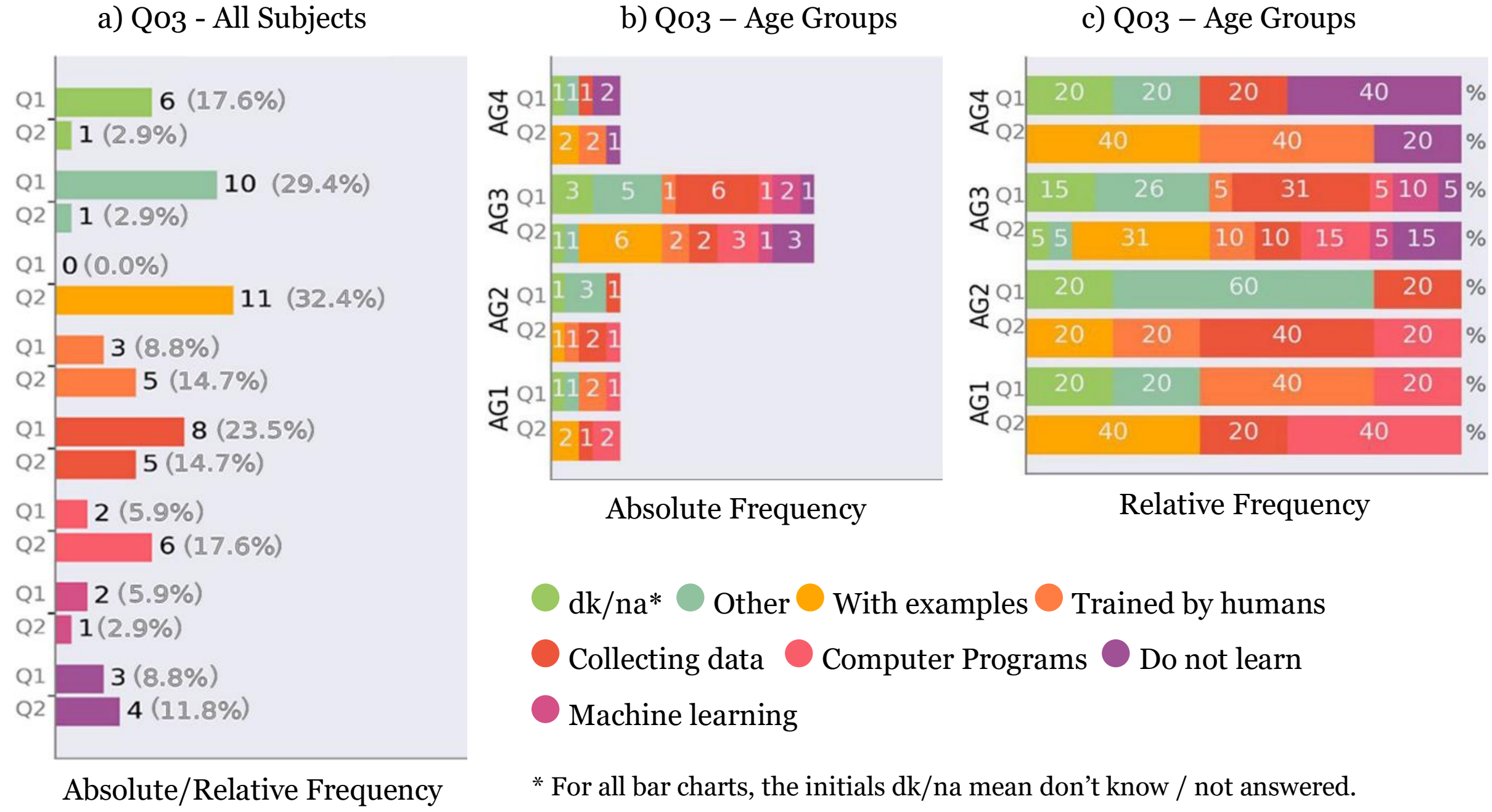

**Fig. 3** Bar charts for Question 3 answers. **a** Pre-posttest all subjects' chart; **b** Pre-posttest age group chart (Absolute Frequency); **c** Pre-posttest age group chart (Relative Frequency)

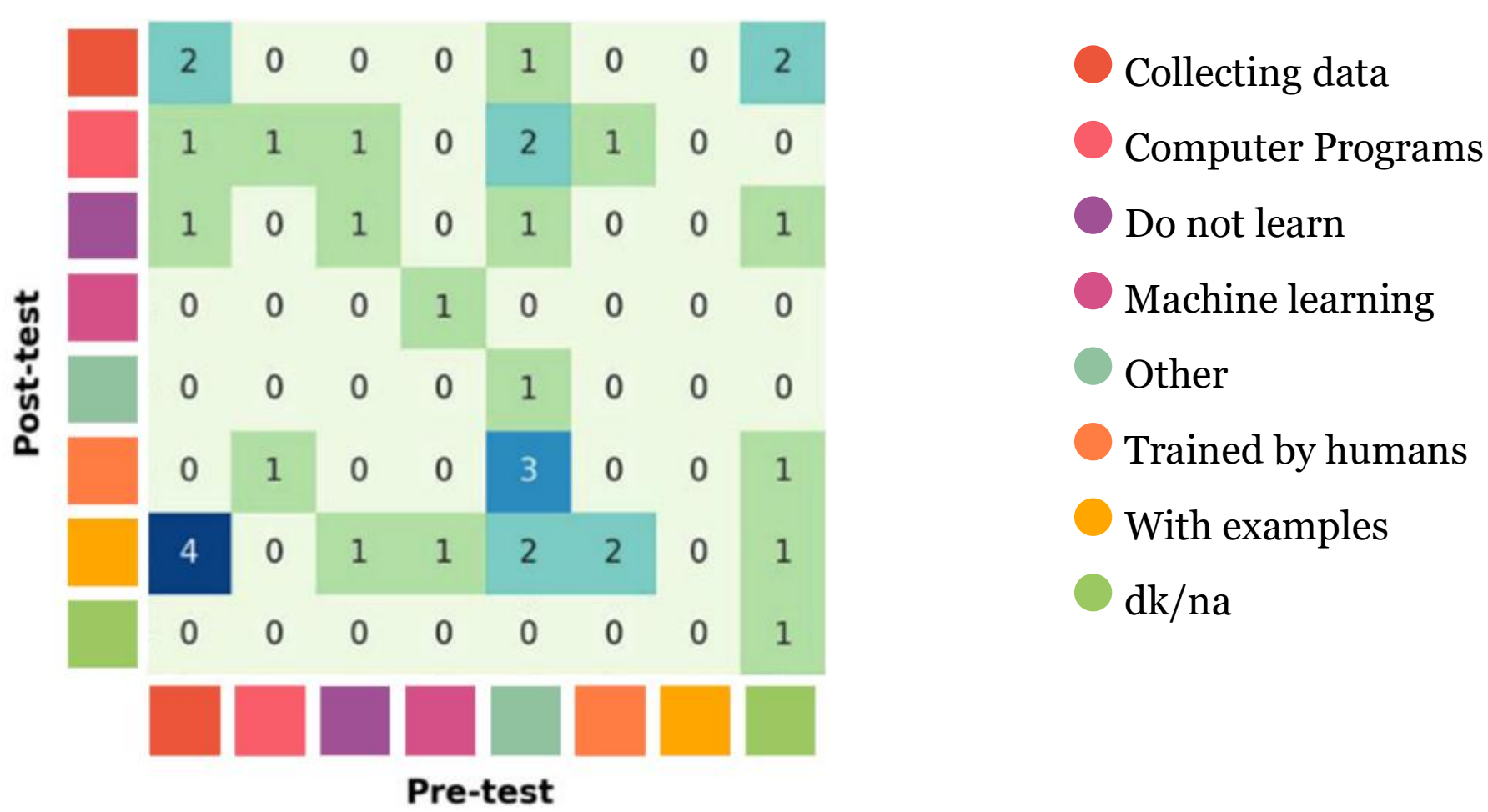

**Fig. 4** Q03-Pre-posttest Matrix

The Pre-Post Test Matrix presented in Figure 4 enables the observation of specific response changes among research participants. For instance, four participants who initially responded with "Collecting data" in the pre-test modified their answer to "With examples" in the post-test. Additionally, the matrix reveals the distribution of responses provided in the post-test by the 11 participants who had initially answered "Other" in the pre-test, as well as by the 6 participants who had either not responded or indicated a lack of knowledge regarding this question prior to participating in the course.

**Q04-Pre-posttest:** If you believe that the computer can recognize an image it has never "seen" but is similar to other photos it has "seen," tell me: how do you think it can do that? (Categorized)

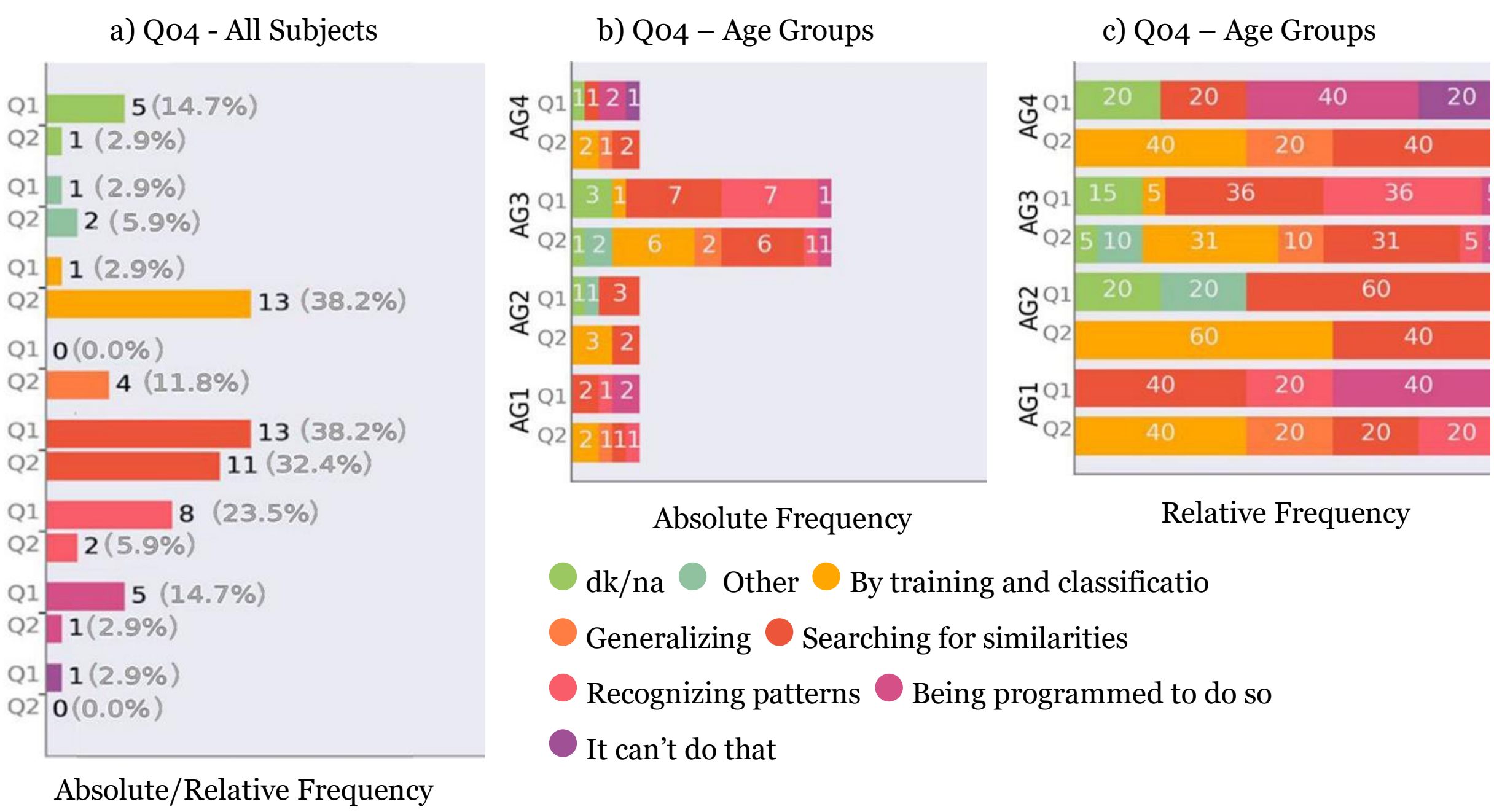

**Fig. 5** Bar charts for Question 4 answers. a Pre-posttest all subjects' chart; b Pre-posttest age group chart (Absolute Frequency); c Pre-posttest age group chart (Relative Frequency)

The bar chart in Figure 5a indicates that approximately 44% of participants (35.3% + 8.8%) who did not verbalize concepts related to training, classification, and generalization for classifying new observations in the pre-test articulated these concepts in the post-test. Furthermore, the stacked bar charts in Figure 5b and Figure 5c demonstrate that this shift in responses occurred consistently across all age groups.

***Q07-Pre-posttest**: What is required to "transform" a conventional computer program into a program capable of learning? (Categorized)*

A comparison of the pre-test and post-test responses to Q07 (as shown in Figure 6) reveals that numerous participants completed the course by providing answers aligned with the material presented in the AIcon2abs course. Ideas such as "Include a learning mechanism" and "Train with examples," which were absent in the pre-test responses, were verbalized by many participants in the post-test.

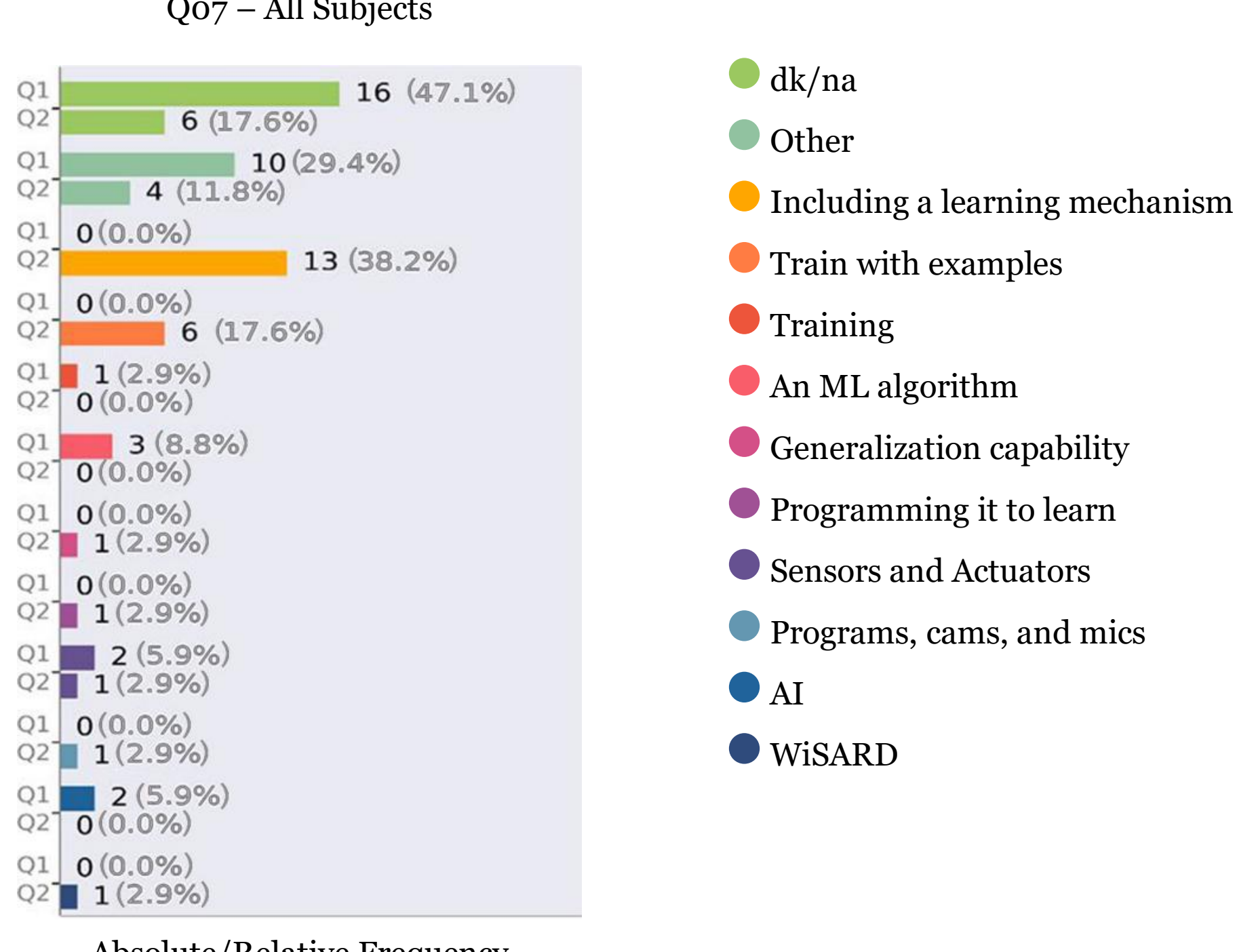

**Fig. 6** Q07-Pre-posttest all subjects' chart

***Q10-Pre-posttest:*** *What is machine learning? (Categorized)*

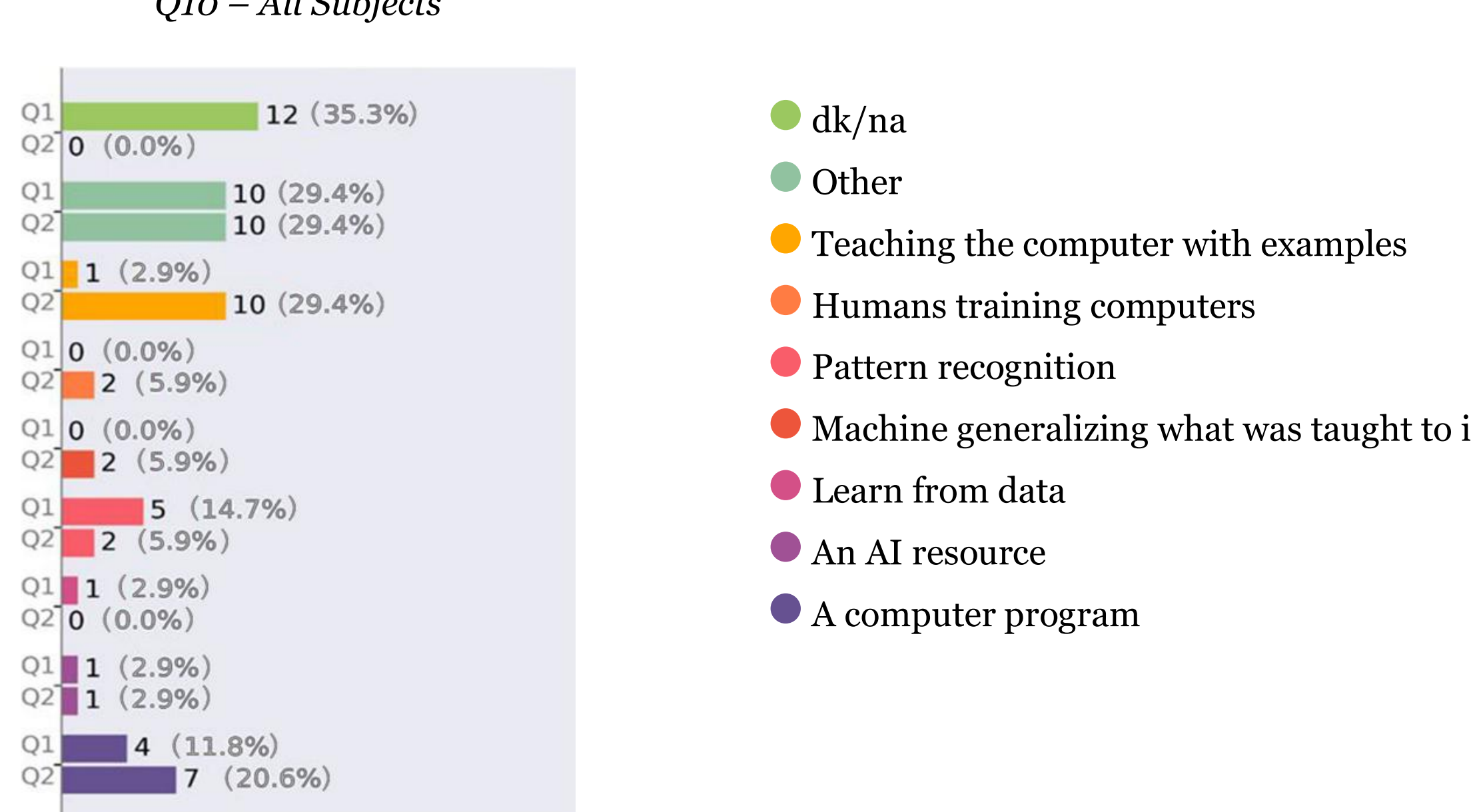

**Fig. 7** Q10-Pre-posttest all subjects' charts

The analysis of the bar chart related to responses to question Q10 (Figure 7) reveals that approximately 32% of participants (26.5% + 5.9%) who did not initially articulate the concepts of "Teaching the computer with examples" or "Humans training computers" as definitions of machine learning in the pre-test expressed these ideas in the post-test. Additionally, all participants who were unsure or did not respond to the pre-test question (35.3%) provided answers in the post-test. Notably, eleven participants (approximately 32% of the total) began the course by offering responses already consistent with the course material.

The Pre-Posttest Matrix in Figure 8 further indicates that among those participants who did not answer or stated they did not know the answer in the pre-test, three changed their responses to "Teaching the computer with examples," one to "Pattern recognition," two to "Humans training computers," two to "A computer program," while two did not incorporate any of the machine learning concepts covered in the course into their post-test responses. Additionally, of the twelve individuals who provided alternative responses in the pre-test, one adjusted their answer to "The machine generalizing what was taught to it," five to "Teaching the computer with examples," and two to "A computer program." Four participants concluded the course without demonstrating clear concepts of machine learning in their responses.

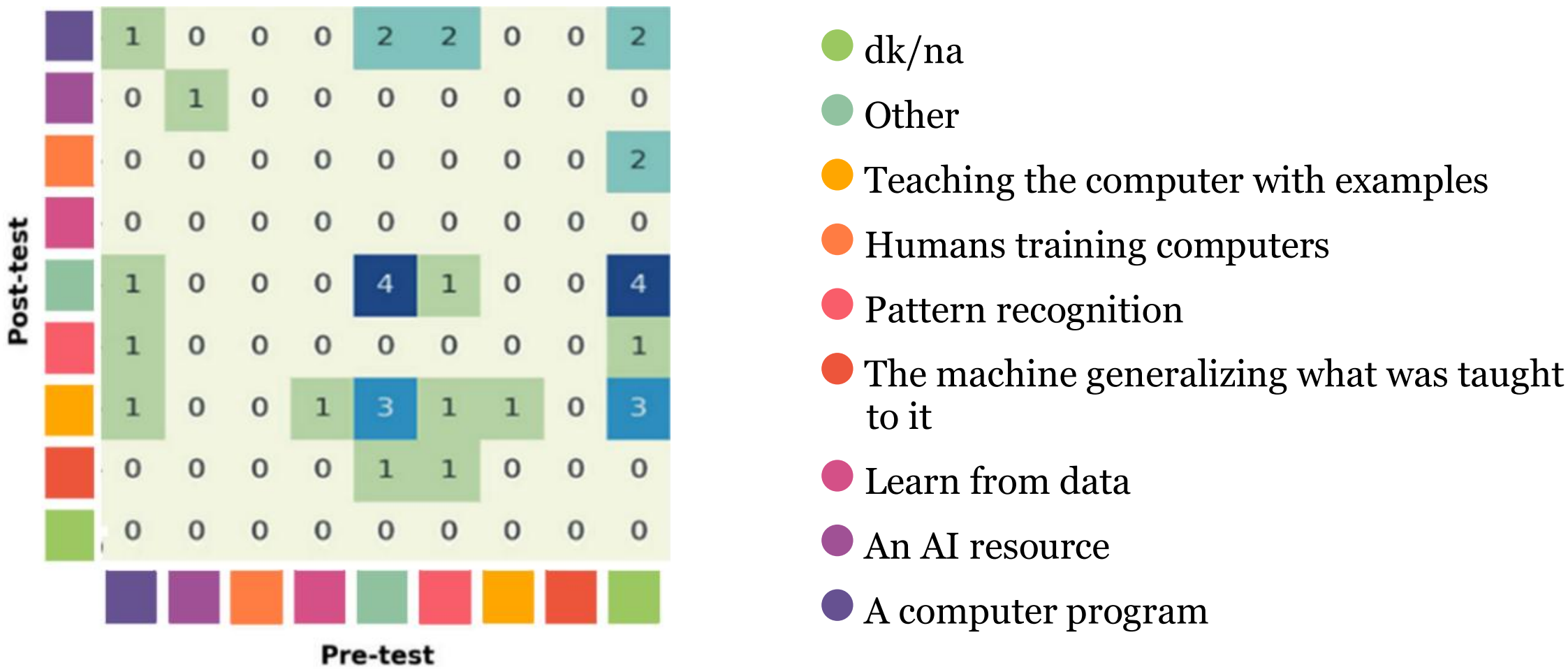

**Fig. 8** Q10-Pre-posttest Matrix

***Q14-Pos-test:*** *Has your understanding of machine learning changed after you took the course? (Categorical) / Q15-Pos-test: What has changed? (Open)*

The bar chart displaying responses to Question 14 (Figure 9) suggests that the course contributed significantly to altering the participants' understanding of machine learning. Illustrative responses to the open-ended question “What has changed in your understanding of machine learning?” (Q15) include: (i) “I now understand how this learning process functions internally,” (ii) “I have learned that machines acquire knowledge through examples,” (iii) “I realized that machines can be continuously trained and improve their accuracy over time,” and (iv) “I previously believed there was only one approach to learning”.

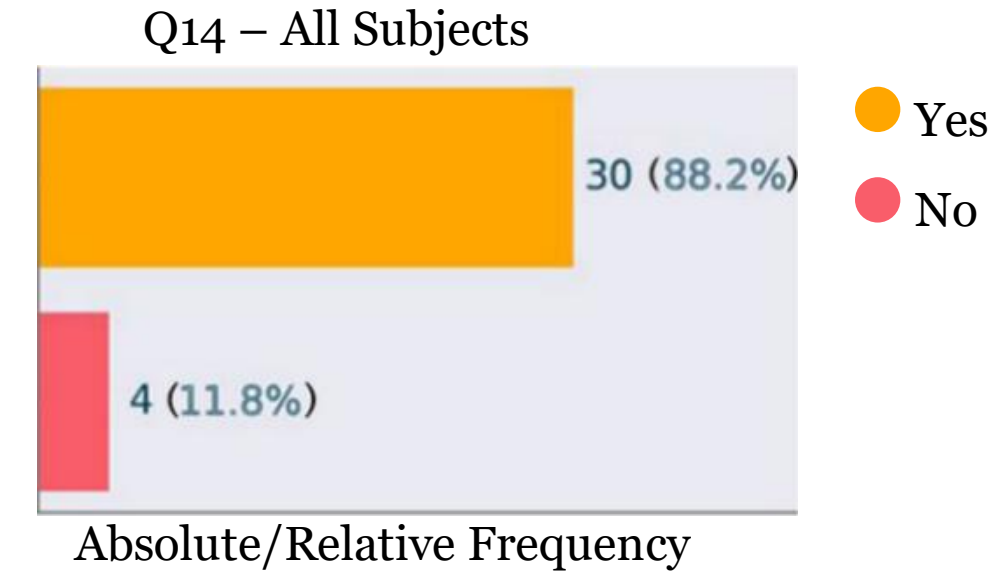

**Fig. 9** Q14-Pos-test all subjects’ charts

***Q18-Pos-test:*** Did you understand the programs made with BlockWiSARD Been 1 = "I did not understand" and 5 = "I totally understand" how would you rate your understanding of the programs? (Ordinal) / Q20-Posttest: Did you understand how WiSARD works by doing the Scratch [interactive ludic] activities? Choose rate from 1 to 5, where 1 = "I cannot understand" and 5 = "I fully understand". (Ordinal)

Bar chart in Figure 10 shows that most of the research subjects could understand the programming challenges performed with BlockWiSARD. Regarding the Scratch ludic activities about the WiSARD model, the results indicate a good understanding of most participants. As shown in Figure 11, scores 4 and 5 represent m16ost of the scores used to evaluate the method.

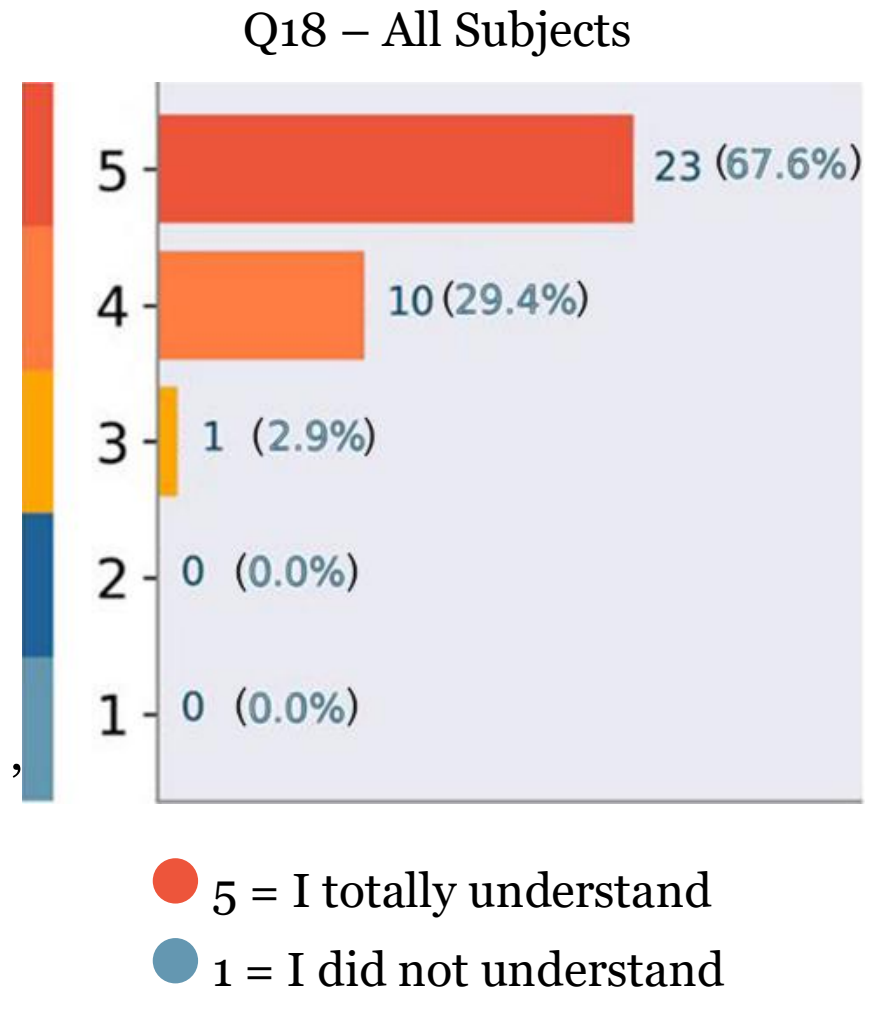

**Fig. 10** Q18-Pos-test all subjects' chart

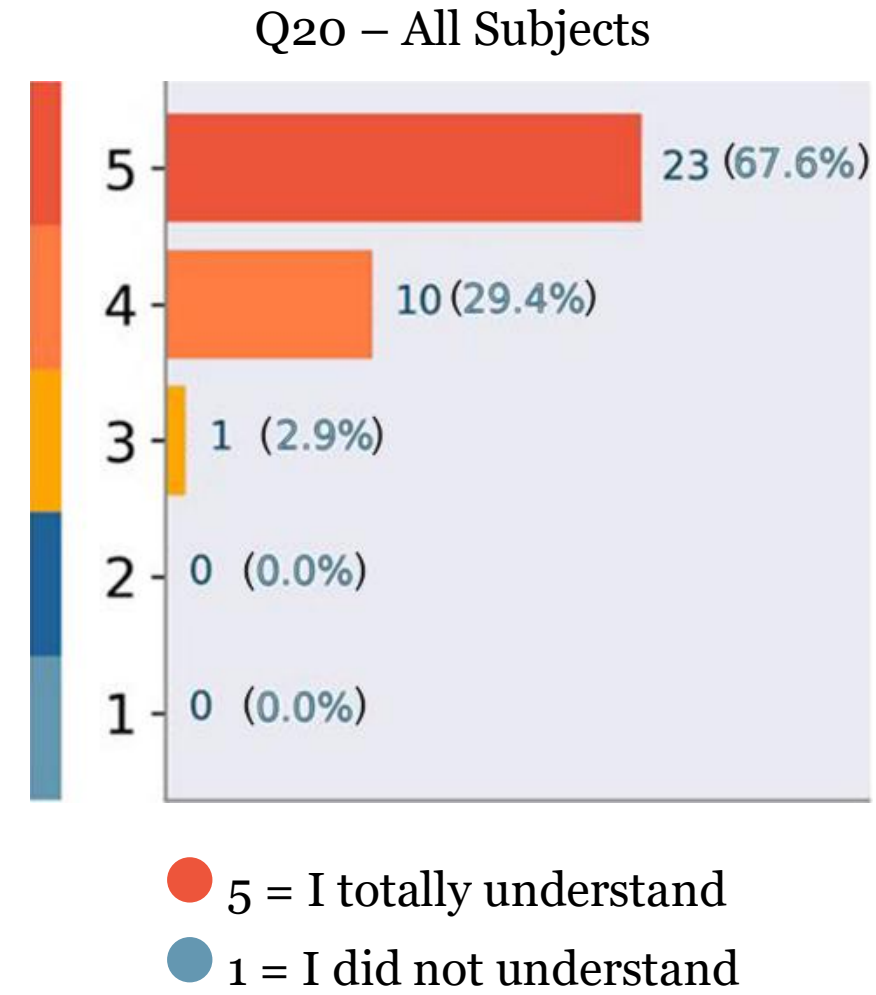

**Fig. 11** Q20 Pos-test all subjects' chart

Once the empirical evaluation was conducted remotely (Due to the COVID-19 pandemic), the IU utilized in the method evaluation did not include the AIcon2abs robotics materials. However, Alpha[15] tests were performed with these materials, as shown in the links below

1. *A Raspberry Pi wheeled robot that can sense, learn, and act (https://youtu.be/KfEOnvSr_kU)*
2. *A wheeled robot (made from simple robotics) sensing, learning, and acting through a program developed with BlockWiSARD (https://youtu.be/GK6GHzuix10)*
3. *BlockWiSARD working alongside an Arduino board (https://youtu.be/xOLXF5LvziQ)*

### 4.1.2. **Hypothesis testing (Inferential analysis)**

Beyond exploratory data analysis, **non-parametric hypothesis tests** (Siegel, 1957) were conducted on the responses to pre- and post-test questions Q03, Q04, Q07, and Q10. Two non-parametric tests were utilized: the **Wilcoxon signed-rank test** and the **Sign test** (Cohen et al., 2002; Kirk, 1978; Pett, 2015; Siegel, 1957). The hypothesis testing described in this section was designed to evaluate the following hypotheses:

[15] "In-house testing performed by the test team (and possibly other interested, friendly insiders)." (Kaner et al., 2008, p.34)

- ***H0*** (**Null Hypothesis**): The AIcon2abs course had no effect on participants' understanding of the content covered regarding machine learning concepts and processes.
- ***H1*** (**Alternative hypothesis**): The AIcon2abs course had a positive effect on participants' understanding of the content covered regarding machine learning concepts and processes.

To conduct these hypothesis tests, scores were assigned to each response category defined for the pre- and post-test questions Q03, Q04, Q07, and Q10 (see introductory paragraphs of Section 4). Table 3 presents the scoring scale applied to attribute scalar values to the four categorized questions related to machine learning concepts and processes addressed in the AIcon2abs course. All hypothesis testing procedures were carried out using IBM® SPSS® Statistics [17].

**Table 3** Criteria adopted to score the categorized answers

| Score | Answer's coherence level regarding the content covered in the course |
|---|---|
| 5 | Answer consistent with the course content |
| 4 | Answer relatively consistent with the course content |
| 3 | Answer slightly consistent with the course content |
| 2 | Answer with little consistency in relation to the course content |
| 1 | Answer not consistent with the course content |
| 0 | Answer equal to “Do not know” or “Blank answer” |

### *Hypothesis test part 1*

For the Wilcoxon signed-rank test to yield robust and reliable results, a primary requirement is that the distribution of the differences between the two dependent samples should be approximately symmetrical. This condition was met between the two dependent samples when considering the total score across all four questions for all 34 research participants. Table 4 and Table 5 display the results of the Wilcoxon signed-rank test and the Sign test for questions Q03, Q04, Q07, and Q10. As indicated, both tests produced identical results.

**Table 4** Wilcoxon signed-rank test results for Q03+Q04+Q07+Q10

| Scores | Significance | Decision | N | (+) dif | (-) dif | Tais |
|---|---|---|---|---|---|---|
| Q03 + Q04 + Q07 + Q10 | <,001. | *Reject H0* | 34 | 28 | 2 | 4 |

**Table 5** Sign test results for Q03+Q04+Q07+Q10

| Scores | Significance | Decision | N | (+) dif | (-) dif | Tais |
|---|---|---|---|---|---|---|
| Q03 + Q04 + Q07 + Q10 | <,001. | *Reject H0* | 34 | 28 | 2 | 4 |

[17] https://www.ibm.com/spss

### Hypothesis test part 2

The Wilcoxon signed-rank test assumption of symmetry was not met when analyzing the scores for each of the four questions individually (even without dividing the data by age groups). Consequently, only the Sign test was applied for these hypothesis tests. Table 6 presents the Sing tests results.

**Table 6** Sign test results for Q03, Q04, Q07, and Q10 (separately) regarding the pre-posttest scores of the 34 research subjects

| Scores | Significance | Decision | N | (+) dif | (-) dif | Tais |
|---|---|---|---|---|---|---|
| Q03 | ,002 | *Reject H0* | 34 | 20 | 4 | 10 |
| Q04 | <,001 | *Reject H0* | 34 | 20 | 3 | 11 |
| Q07 | <,001 | *Reject H0* | 34 | 19 | 2 | 13 |
| Q10 | ,002 | *Reject H0* | 34 | 22 | 5 | 7 |

The p-value for statistical significance was set at 0.05 when defining test parameters in IBM® SPSS® Statistics. A p-value of 0.05 is a standard threshold in hypothesis testing, indicating a 5% probability that the observed results could occur purely by chance. However, in most cases, the statistical significance calculated by the tests yielded a p-value below 0.001. This level of significance implies that the likelihood of the observed results occurring by chance is approximately 1 in 1,000, or 0.1%.

**The results obtained from the hypothesis tests (leading to a rejection of H0) support the positive effect of the AIcon2abs course on the research participants.** In addition to statistical significance, the marked prevalence of positive post-test outcomes compared to pre-test results is also noteworthy.

## 4.2. Data Analysis from a phenomenological perspective

This study also employed a phenomenological analysis approach. As outlined in Section 2, we adopted the phenomenological methodology proposed by Creswell (2016), which involves describing the phenomenon experienced by research subjects and the context in which this experience unfolded. Additionally, in a few comprehensive paragraphs, the researcher presents a portrayal of the essence of participants' perceptions of their lived experience. This narrative is constructed by reducing and organizing participants' testimonies into significant quotations classified by thematic categories.

### 4.2.1. The lived experience and its context

Section 3 detailed the context in which participants experienced the observed phenomenon: an online course conducted as part of research on a method intended to provide insight into the inner workings of machine learning and AI. Participation was voluntary, and motivations varied, including (i) curiosity and interest in the topics covered (computer programming, AI, and machine learning), (ii) the desire to expand knowledge, and (iii) the opportunity to contribute to scientific research.

In Subsection 4.2.2, titled *The AIcon2abs Course*, we present a narrative capturing the essence of participants' experiences. This narrative synthesizes perceptions derived from responses to post-test questions 23, 25, and 26 (Table 1), alongside additional comments on the learning platform. To develop the narrative, we organized participants' responses into four categories: (i) Feelings, (ii) Didactics and Content, (iii) Learning, and (iv) Reflections. The thematic categorization and original statements in Portuguese are available in AIcon2abs (2023).

4.2.2. The AIcon2abs course

The following narrative serves as a composite testimonial reflecting the common feelings, perceptions, and reflections of the 34 participants regarding the AIcon2abs course:

*"I chose to enroll in the AIcon2abs course out of curiosity, hoping to learn more about artificial intelligence and machine learning. The course was offered online, featuring pre-recorded lessons, and began with the story of Zrow, a robot from another planet on a quest to learn about Earth. The narrative of Zrow's adventures provided a stimulating framework for the course, evoking the dreams of my childhood. This method of presenting the course content was highly engaging. The course materials were of exceptional quality, both technically and pedagogically. The content was intuitive and well-explained, requiring no prior knowledge. This shows that complex topics, like AI and machine learning, can be made accessible to a broad audience. The themes were explored through interactive, simple, and creative activities, including Scratch-based games and animations, which were very engaging. The course was designed to accommodate various age groups, with thoughtful consideration for older adults and children alike.*

*Throughout the course, I gained insights into programming, machine learning, and AI. I learned that machines acquire knowledge through examples and that they operate based on programmed instructions. I also learned about WiSARD, which was surprisingly easy to understand. I gained a sense of the processes that allow machines to 'learn' and discovered different methods for training AI. This course opened a new world to me, a realm in which I had been an outsider. Some activities, however, could benefit from additional time and examples. I also believe that face-to-face interaction could enhance learning, particularly for younger participants who may find even brief video content tiring.*

*The course prompted me to reflect on the need for greater awareness of AI. The general public should understand AI to participate in discussions about its future. This is knowledge that should permeate our schools and academic institutions. The course raised several questions: What applications of AI influence our daily lives without our awareness? Will most people remain uninformed, leaving knowledge holders to shape decisions for us? I reflected on the implications of AI for free will and its potential for transforming our understanding of autonomy. This experience broadened my understanding of the world and affirmed the remarkable capacity of the human mind. Ultimately, I found the course to be a truly rewarding experience."*

### 4.3. Discussion

This section discusses the results through the lens of the DSR method as outlined by Dresch et al. (2015) and Pimentel et al. (2019), which guided this research (see Figure 1). We will first evaluate the three primary assessments expected in DSR-based research per Pimentel et al. (2019). Next, **we will formalize the construction and contingency heuristics**, the core contributions of the DSR method as proposed by Dresch et al. (2015). Finally, we will address some limitations of this study.

#### 4.3.1. **Is the artifact valid?**

Results in Subsections 4.1 and 4.2 indicate that (i) the research objectives (Subsection 1.2) were satisfactorily achieved, (ii) the requirements (Subsection 1.3) were met, and (iii) the anticipated performance (Subsection 1.5) was realized. **These outcomes support the conclusion that the AIcon2abs method is valid**, meeting the requirements of Evaluation 1 in the DSR map presented in Section 2.

#### 4.3.2. **Was the problem solved?**

The data indicate that most participants found the AIcon2abs course activities easy and that many developed a foundational understanding of machine learning in its most basic form. Furthermore, participants' satisfaction with the course suggests that learning through AIcon2abs was a rewarding experience. According to Simon (1996), consensus among participants serves as a criterion for determining whether a solution is satisfactory. **The data presented here show that participants were generally satisfied, indicating that AIcon2abs is a suitable method for addressing its intended problem**.

#### 4.3.3. **Do the theoretical conjectures seem to be valid?**

In Subsection 1.4, we outlined six theoretical conjectures guiding the development of the AIcon2abs method. The findings suggest the validity of these conjectures, **particularly the central one, which posits that *learning abstract concepts from computer science can be facilitated through practices that highlight observable aspects of these concepts in the concrete world.*** It is valid to mention that none of the research or approaches found in the Queiroz et al. (2021) literature review had the same target audience as the AIcon2abs method: the general public. Those studies were exclusively dedicated to a specific age group: (i) children, (ii) adolescents, or (iii) undergraduate course students. Therefore, **it is worth stating that this study revealed that this theoretical conjecture seems valid for a diverse audience of different ages and backgrounds (including K–12 students)**.

Regarding conjecture 5 (Subsection 1.4), after implementing the AIcon2abs course, we realized that the agent-based approach could be enhanced by integrating it with the *Input-Process-Output* (IPO) model[18]. This combination could further facilitate understanding the fundamental logic of machine learning in its most basic form.

#### 4.3.4. **AIcon2abs construction heuristics**

*Construction heuristics* establish rules for the internal environment of the artifact and for its construction. When formalized, these heuristics can aid other researchers in developing artifacts to address similar issues. Construction heuristics represent a core contribution to knowledge within the DSR framework (Dresch et al., 2015; Gregor & Jones, 2007; Simon, 1996; Venable, 2006). Fig.12 presents the formalized construction heuristic derived from the AIcon2abs method's design, development, implementation, and evaluation.

---

[18] A computer works based on a cycle defined as Input-Process-Output (IPO). This is an elementary concept widely used in introductory computing and can be summarized as: *Input (Anything that is put into a system)* - > *Process (The steps or actions taken by the system)* -> *Output (The result of the process)* (Maharaj et al., 2023, p.147).

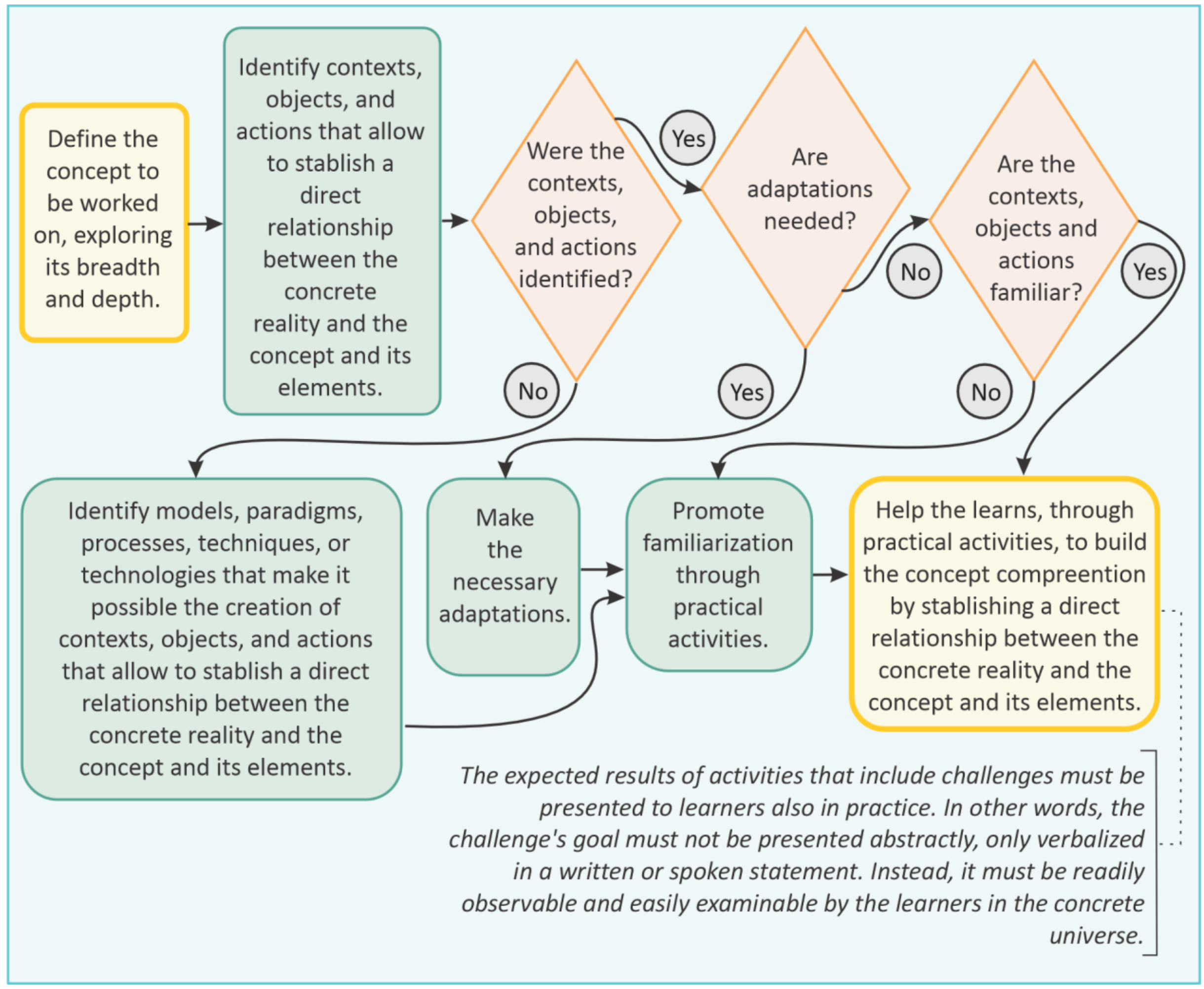

**Fig. 12** AIcon2abs construction heuristic

### 4.3.5. **AIcon2abs contingency heuristics**

**Contingency heuristics specify, when applicable, the adjustments required to adapt an artifact for effective functioning in different external environments** (Dresch et al., 2015). For the AIcon2abs method, the external environment is defined as its target audience. During data analysis, some participants highlighted certain aspects of the AIcon2abs method that made it difficult to understand some specific content. Notably, these challenges were exclusively reported by adult participants. The most frequently cited issues included:

- Some content was perceived as overly "complex," requiring more than a single engagement with the related activity for adequate comprehension.
- Activities associated with more complex content had insufficient examples to support understanding.
- The recommended course duration was inadequate for consolidating certain content.

These findings represent contingencies that should be addressed to meet the learning objectives of the AIcon2abs method, particularly when applied to an audience with the specified learning needs. The identified contingencies focus on the difficulty of adequately understanding some themes worked on in the AIcon2abs IU. In alignment with the *Differentiated Learning educational approach* (Tomlinson, 1999), we propose the contingency heuristic illustrated in Figure 13. This heuristic is founded on established theoretical principles and pedagogical practices. However, a new Design Science Research (DSR) cycle, encompassing steps from Artifact Design to Evaluation (see Figure 1), is recommended to validate its efficacy.

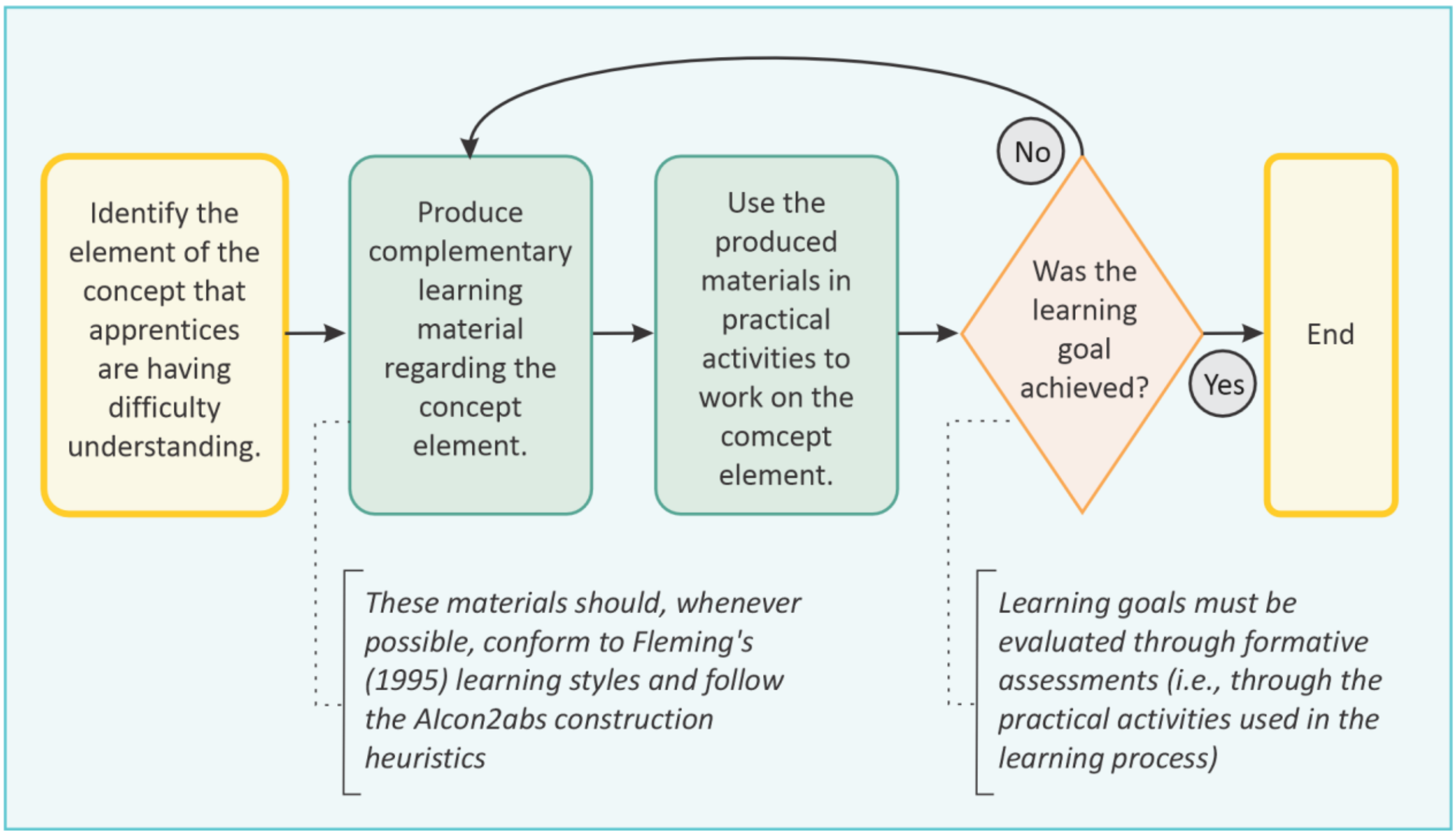

**Fig. 13** AIcon2abs contingency heuristic

### 4.3.6. **Limitations of the study**

While efforts were made to address threats to the validity of this research, some limitations remain. Regarding **external validity**, despite the promising results observed for the AIcon2abs method within the context of this study, its effectiveness in other settings is not assured. The method was evaluated based on a relatively small sample of 34 participants. While the sample exhibited diversity in certain respects, it was specific in others: (i) all adults had either completed or were attending higher education, and (ii) all children and adolescents had access to quality education. Concerning **internal validity**, data were not collected from participants who withdrew from the AIcon2abs course, precluding comparisons between those who completed the course and those who did not. Furthermore, all participants were presumed to be prepared to engage with the course content, potentially influencing the outcomes. These considerations highlight the need for further evaluations of the AIcon2abs method across diverse contexts. Suggestions for future research and potential directions are presented in Subsection 5.2.

## 5. Conclusion

### 5.1. Concluding summary

This study aimed to address the question: ***Is AIcon2abs a suitable method to support a diverse audience of different ages and backgrounds in understanding the concept of machine learning in its most basic form in an uncomplicated and satisfactory way?*** The results presented here affirmatively answer this question. Moreover, they demonstrate that the AIcon2abs method effectively achieved its objectives by incorporating the following four strategies, which constitute the primary contribution of this research:

- Employing Active Learning strategies.
- Presenting the fundamental concept of machine learning (in its most basic form) through concrete examples that are familiar to the target audience and aligned with the adopted machine learning model constructs.
- Integrating the learning algorithm instantiation, along with training and classification primitives, as basic commands within a conventional programming structure.
- Utilizing a machine learning algorithm whose training and classification processes are readily observable and replicable, allowing learners to observe the internal actions of the machine during the learning process.

Additionally, **the formalization of construction and contingency heuristics** for the AIcon2abs method (outlined in Subsections 4.3.4 and 4.3.5) constitutes a valuable contribution of this study. Other notable contributions include: (i) the **Pre-Posttest Matrix data visualization tool**, an adapted use and interpretation of the Confusion Matrix (Section 4), and (ii) **the blending of the Design Science Research (DSR) method by Dresch et al. (2015) with elements from the DSR map proposed by Pimentel et al. (2019)** (Section 2).

The AIcon2abs components were highly rated by nearly all participants—children, adolescents, and adults—who took part in the empirical evaluation. Having been rigorously evaluated and validated**, the AIcon2abs method is now available to those interested in demystifying artificial intelligence, particularly machine learning, to the general public.**

### 5.2. Future work

To further enhance the AIcon2abs method and broaden its applicability to include new audiences, the following future work would be of significant value:

- Develop English versions of the instructional videos and playful activities using the WiSARD model.
- Incorporate the Input-Process-Output (IPO) concept alongside the Intelligent Agents concept in the design of the AIcon2abs instructional unit, to reinforce understanding of the fundamental differences between traditional computer systems and those utilizing AI mechanisms.
- Produce the additional learning materials proposed in the formalized contingency heuristics.
- Offer a face-to-face version of the AIcon2abs course to: (i) apply the contingency heuristic developed in this study and (ii) incorporate educational robotics materials in the BlockWiSARD activities.
- Adapt the AIcon2abs method for illiterate learners.
- Investigate the feasibility of adapting the AIcon2abs method for individuals with disabilities, such as the visually impaired and the deaf.
- Explore the potential of using AIcon2abs construction heuristics to develop artifacts introducing other AI paradigms to the general public.
- Apply the AIcon2abs construction heuristics to design materials addressing other complex computing concepts, such as Object-Oriented Programming Paradigm.

## Ethics statement

This research has been approved by the CEP/HUCFF/FM/UFRJ[19] human research ethics committee: CAAE: 44444621.7.0000.5257, approval date: June 22, 2021, approval order number: 4.861.208, 2021. All adults, parents, or legal guardians gave their informed consent, and all minors gave their informed assent before participating in this study.

## Conflict of interest statement

The authors declare that they have no known competing financial interests or personal relationships that could have appeared to influence the work reported in this paper.

## Data availability

The datasets used for the data analysis presented in this paper are available at AIcon2abs (2023).

## Acknowledgements

This study was financed in part by the Coordenação de Aperfeiçoamento de Pessoal de Nível Superior-Brasil (CAPES)-Finance Code 001.

## Funding

Coordenação de Aperfeiçoamento de Pessoal de Nível Superior-Brasil (CAPES)- Finance Code 001. (*Graduate Fellowship*).

[19] *https://www.hucff.ufrj.br/comite-de-etica-em-pesquisa/*